\documentclass{vgtc}                          

\ifpdf
  \pdfoutput=1\relax                   
  \usepackage{graphicx}                
  \DeclareGraphicsExtensions{.pdf,.png,.jpg,.jpeg} 
\else
  \usepackage{graphicx}                
  \DeclareGraphicsExtensions{.eps}     
\fi%

\graphicspath{{figures/}{pictures/}{images/}{./}} 

\usepackage{microtype}                 
\PassOptionsToPackage{warn}{textcomp}  
\usepackage{textcomp}                  
\usepackage{mathptmx}                  
\usepackage{times}                     
\usepackage{cite}                      
\usepackage{tabu}                      
\usepackage{booktabs}                  

\onlineid{1014}

\vgtccategory{Paper Track}

\vgtcinsertpkg

\usepackage{amsfonts}

\title{Uncertainty-aware Topic Modeling Visualization}

\author{Valerie M\"uller\thanks{Institute of Computer Science, e-mail: linsen@uni-muenster.de} \and Christian Sieg\thanks{Institute for German Philology, e-mail: christian.sieg@uni-muenster.de} \and Lars Linsen$^*$
}
\affiliation{\scriptsize Westf\"alische Wilhelms-Universit\"at M\"unster, Germany}

\abstract{Topic modeling is a state-of-the-art technique for analyzing text corpora. It uses a statistical model, most commonly Latent Dirichlet Allocation (LDA), to discover abstract topics that occur in the document collection. However, the LDA-based topic modeling procedure is based on a randomly selected initial configuration as well as a number of parameter values than need to be chosen. This induces uncertainties on the topic modeling results, and visualization methods should convey these uncertainties during the analysis process. We propose a visual uncertainty-aware topic modeling analysis. We capture the uncertainty by computing topic modeling ensembles and propose measures for estimating topic modeling uncertainty from the ensemble. Then, we propose to enhance state-of-the-art topic modeling visualization methods to convey the uncertainty in the topic modeling process. We visualize the entire ensemble of topic modeling results at different levels for topic and document analysis. 
We apply our visualization methods to a text corpus to document the impact of uncertainty on the analysis.
} 

{\begin{itemize}\setlength{\itemsep}{-0.5pt}\setlength{\parsep}{-0.5pt}}%
{\end{itemize}}
{\begin{enumerate}\setlength{\itemsep}{-0.5pt}\setlength{\parsep}{-0.5pt}}%
{\end{enumerate}}

\begin{document}


\firstsection{Introduction}

\maketitle

Analyzing corpora of unstructured text data is a main challenge in many application areas such as Digital Humanities and Social Sciences or Medicine and Health Care. Corpora are either directly generated in digital form or being digitalized by scanning the documents and applying optical character recognition tools. Given the digital corpus, a powerful and widely used approach for discovering semantic structures is \textit{topic modeling}. It uses statistical models such as \textit{Latent Dirichlet Allocation} (LDA) to detect abstract \textit{topics} based on the frequency of occurrences of (key) words or \textit{terms}.

During the analysis of a topic-modeling result, \textit{multiple facets} are investigated such as which terms relate to which topic, which topics relate to which documents of the corpus, and which terms relate to which documents. Interactive visual analysis tools can support such an analysis. However, existing analysis tools operate on one topic-modeling result, while there are several aspects that make the topic models \textit{uncertain}. Topic modeling such as LDA-based approaches typically start with some random initial configuration and update the configuration during an optimization process. The optima are local though and depend on the initial configuration, which makes the outcome uncertain. Moreover, the topic modeling process is steered by a number of parameters that may severely influence the outcome. Most prominently, one typically has to choose the number of topics as an input parameter, which is usually unknown a priori. The choice of these parameters also render the outcome uncertain.

We present an uncertainty-aware interactive visual analysis of unstructured text corpora based on topic modeling. We capture the uncertainty in the topic modeling outcome by generating ensembles of topic models. We further propose two measures for estimating the topic modeling uncertainty within the ensemble. Our visualization tool then uses multiple coordinated views to investigate the multiple facets of the topic modeling ensemble, which allows for an analysis of the topic modeling outcome while revealing its uncertainty. 

The individual contributions of our work can be summarized as follows:
    (1) Generation of topic modeling ensembles to capture uncertainties in  topic models.
    (2) Uncertainty measures for topic models.
    (3) Uncertainty-aware interactive visual analysis tool for multiple facets of topic models.
    (4) Analysis of the impact of sources of uncertainties on the topic models.
    (5) Application to uncertainty-aware semantic analysis of text corpora.


\section{Related Work}
\label{sec:relatedwork}

\noindent
\textbf{Visualization of Text Corpora.}
J\"anicke et al.~\cite{jaenicke} present a survey on visual text analysis in Digital Humanities, where they develop a task-based taxonomy. 
Zhang et al.~\cite{zhang} survey visualization approach for topic analyses. The approaches are categorized into visualization of topic content, topic relationship, and topic evolution. 
Our work relates to  topic relationship and topic content. Topic evolution is beyond the scope of this work. 

\textit{Topic relationship visualization} relates detected topics to each other. In case the topics are captured in a hierarchy, hierarchical representations \cite{dou} can be used. 
Otherwise, the similarity of topics is often reflected by a 2D layout design, e.g., by using graph layouts \cite{topicNets,iVisClustering, wang} or embeddings \cite{ji,LDAvis}. The latter can also be encoded using a map metaphor \cite{gansner}. When topics are obtained by topic modeling, what has been largely neglected is the sensitivity of the outcome to certain choices made during the modeling procedure. We propose to capture the uncertainty in the topic modeling process and to analyze its robustness by investigating topic modeling ensembles. Alexander and Gleicher~\cite{alexander} took a first step in this direction by comparing individual topic models. Similarly, El-Assady et al.~\cite{El-Assady2018} propose to compare topics models to each other and define topic matching based on document overlap and keyword similarity. Three topic similarities using information theory (Jenson-Shannon divergence, a smoothed and a symmetric Kullback-Leibler divergence) were defined Within a purely analytical approach~\cite{Niekler2012}.  We capture and visualize the uncertainty in topic modeling, which also requires the definition of topic similarities.

\textit{Topic content visualization} can be performed by relating topics to terms, e.g., by visualizing frequency of term occurrences per topic via a term-topic matrix \cite{chuang}. Bar charts are also widely used to visualize term frequencies \cite{chuang,LDAvis,topicFlow}. Another common approach to visualize the content of a topic is to visually represent the frequencies of the most popular terms using word clouds \cite{heimerl}.
In addition, topics can be related to documents within the document collection such as a text corpus \cite{chuang,TMVisBlei2012}. Serendip~\cite{serendip} is a comprehensive visual analysis tool that allows for relating topic content visualization with meta-data analysis and highlighted textual visualization. We build upon these existing approaches, combine them with topic relationship visualizations, and enhance the entire approach towards an uncertainty-aware topic  analysis.  

\medskip
\noindent
\textbf{Uncertainty Visualization.}
MacEachren~\cite{maceachren:1992:VUI} presented early ideas on uncertainty visualization by discussing the suitability of the visual variables introduced by Bertin~\cite{bertin}. 
A survey by Bonneau et al.~\cite{Bonneau2014UncertaintyVis} identified four types of uncertainty visualization: Comparative investigation of different variants using juxtaposed or superimposed layouts, mapping to an unused visual variable (often color~\cite{Al-Taie2014,Chen2015}), glyphs encodings~\cite{ristovski2}, and mapping to blur or texture. 

The modeling of uncertainty commonly relates to probabilistic models that capture stochastic distributions of uncertain values. 
For specific realizations, the probabilistic model is sampled using some sampling strategy such as Monte Carlo sampling. The set of different realizations then form an \textit{ensemble}, where the ensemble then captures the inherent uncertainty. We follow this idea in Section~\ref{sec:uncertaintymodeling}. 
Given an ensemble, visualization approaches to analyze the ensemble have been generated. Recently, there was a large amount of work on ensemble visualization of spatial simulation data
~\cite{ensemble}. We are dealing with ensembles of topic models though. Most related are ensemble approaches that deal with cluster ensembles~\cite{altaie14,Stehl2003}. Another related approach investigates  stability of topic models within an ensemble via matrix factorization~\cite{Belford2017}.
In terms of visualization of topic modeling ensembles, our work is most closely related to work on stability of dimensionality reduction methods~\cite{Reinbold2019} 
and the analysis of behavior of actors using an LDA ensemble~\cite{Chen2019}. Our goal is very different though.

\section{Background}
\label{sec:background}

\noindent
\textbf{Topic Modeling.}
Topic modeling of unstructured text corpora is a distant reading technique~\cite{moretti2013distant}, where automatic computer-based methods are applied for text data analysis. It is nowadays more and more common that distant-reading methods complement close reading to cope with the large amount of data~\cite{jaenicke}.
Topic models are based on probabilistic models to detect patterns in document collections. 
Input to a topic-modeling approach is the \textit{document-term matrix} that stores for each document of the corpus the number of occurrences of each term.
A topic model then describes which topics contribute how much to which documents and which terms contribute how much to which topics.
Thus, the topic model provides a global overview of the corpus.

Different topic-modeling approaches exist. 
Presumably the most prominent 
approach is  Latent Dirichlet Allocation (LDA), which we will also use.
However, since all topic-modeling approaches use randomness and are steered by input parameters, the general concepts introduced in this paper can be transferred to any topic-modeling approach. 

\medskip
\noindent
\textbf{Latent Dirichlet Allocation.}
LDA is a generative statistical model that explains a set of observations by a set of unobserved groups or clusters.
In the context of topic modeling, the observations are provided in the form of the document-term matrix and the groups are the topics.

During the generative step, the topic-modeling problem is reversed.
Thus, it is assumed that the topics and their distributions are known and that the documents and their terms are latent variables. 
First, the topics are initialized by randomly assigning to each topic terms and respective probabilistic term distributions.
Second, one chooses for each document a probabilistic topic distribution.
Third, one chooses for each term of each document a topic $T$ based on the the document's topic distribution.
Then, one chooses a term from topic $T$ based the topic's term distribution.
This process generates the documents.
The probabilistic distributions for both topics and terms  use Dirichlet distributions. 

After having generated the documents, the process is reversed, i.e., now documents and terms are known and topics and topic distribution are the latent variables, which is referred to as inference. Inference generates topics that describe the given corpus as precisely as possible. 
Optimization algorithms such as Gibbs sampling~\cite{Griffiths2004Infernece} are applied.
We use the LDA implementation of the machine learning for language toolkit (MALLET).
The outcome are topics with respective terms and term distributions as well as documents with respective topics and topic distributions.

\medskip
\noindent
\textbf{Uncertainties in Topic Modeling.}
First, 
topic modeling is a process that involves \textit{randomization}. Thus, the outcome of the topic modeling is different when running the topic-modeling algorithm multiple times, directly inducing uncertainties. The random initial configurations for the topic-modeling process lead to the generation of only locally optimal solutions. 

Second, all commonly used topic-modeling approaches need the user to specify the number of topics $k$~\cite{Croft2006}. This number $k$ is a priori unknown 
when analyzing a new corpus, but is required as an \textit{input parameter}. 

Third, each topic-modeling approach typically comes with a number of method-specific input parameters (aka hyperparameters) that need to be adjusted to the corpus at hand and, thus, induce further uncertainty. 
In the case of LDA, there are two input parameters $\alpha$ and $\beta$, which are parameters of the Dirichlet distributions. Parameter $\alpha$ determines how many topics shall be assigned to a document, and parameter $\beta$ determines how many terms shall be assigned to a topic. The values of both parameters should be small, but the exact choice is a priori unclear. MALLET uses default values $\alpha=\frac5k$ and $\beta=0.01$ and 
also supports their automatic optimization, but optimal solutions cannot be guaranteed. 
(The number of iterations within the LDA algorithm, however, 
just needs to be chosen large enough and was set to 10,000 for our experiments.)


Bonneau et al.~\cite{Bonneau2014UncertaintyVis} categorize uncertainties in visualization into sampling-based uncertainties, model-based uncertainties, and uncertainties in the visualization process. Topic-modeling approaches such as LDA use sampling of probabilistic distributions and models, where hyperparameters need to be set optimally. Hence, we are facing sampling uncertainties as well as model uncertainties.

\section{Uncertainty Modeling}
\label{sec:uncertaintymodeling}


\medskip
\noindent
\textbf{Capturing Uncertainties.}
Both sampling and model uncertainties can be captured using an \textit{ensemble approach}. To capture sampling uncertainties, we simply run the topic-modeling approach with the same parameter settings multiple times. In each run, the randomization step draws samples from the given probabilistic model. This results in a topic modeling ensemble, where sampling uncertainties are reflected by variances in the topic-modeling outcome.
To capture model uncertainties, we run the topic-modeling approach  multiple times but now with different parameter settings (for parameters $\alpha$, $\beta$, or $k$). Thus, the variances in the results capture the model uncertainty associated with the respective parameter.

\medskip
\noindent
\textbf{Topic Similarity.}
Uncertainties are commonly related to measuring variances in observations. 
We want to measure how (dis-)similar two topics of different topic models are to estimate how much the topic models concur, i.e., how certain one can be about the topic.

Topics are given in the form of a probability distribution over all considered terms. \textit{Kullback-Leibler divergence} is a common entropy-based measure for quantifying how much a probability distribution $\mathbf{a}$ differs from a given reference probability distribution $\mathbf{b}$~\cite{Niekler2012}. Given a probability space $X$, it is defined by
\begin{equation}
    D_{KL}(\mathbf{a} \| \mathbf{b}) = \sum_{x\in X} \mathbf{a}(x) \cdot \log \frac{\mathbf{a}(x)}{\mathbf{b}(x)}
    \label{eq:KL}
\end{equation}
However, this definition is not symmetric. Symmetry in measuring topic similarity is desired though. The \textit{Jensen-Shannon divergence} is a related symmetric measure that could be used instead~\cite{Niekler2012}.

Another possibility is to interpret the probability distribution as a multi-dimensional vector and compute a vector (dis-)similarity.  \textit{Euclidean distance} is a commonly applied measure for computing distances in a multi-dimensional Euclidean space. However, one desired characteristic of the LDA approach is that only few terms contribute to a topic. Therefore, the multi-dimensional data are sparse. Euclidean distances on sparse multi-dimensional data tend to be very small. 
Therefore, it is more desirable to use 
the \textit{cosine similarity}, which for two vectors $\mathbf{a}=(a_1,\ldots,a_n)\in \mathbb{R}^n$ and $\mathbf{b}=(b_1,\ldots,b_n) \in \mathbb{R}^n$ is defined as 
\begin{equation}
    S(\mathbf{a},\mathbf{b)}=\frac{\sum_{i=1}^n {a_i\cdot b_i}}{\sqrt{\sum_{i=1}^n {a_i^2}}\cdot\sqrt{\sum_{i=1}^n {b_i^2}}}
    \label{eq:cosine}
\end{equation}

Niekler and J{\"a}hnichen~\cite{Niekler2012} compared Jensen-Shannon divergence to cosine similarity for topic models and concluded that the cosine similarity achieves better results in terms of average deviation of computed to desired similarity values. Consequently, we will use cosine similarity in the following.

\medskip
\noindent
\textbf{Uncertainty Measures.}
We will introduce two uncertainty measures for topic modeling ensembles. The \textit{matching uncertainty} shall measure how (un-)certain a topic of one topic model can be matched with the topic of another topic model. If the similarity of a topic of the first topic model is high to exactly one topic of the second topic model (i.e., low for all the other topics of the second topic model), then we can be quite certain that the same topic was found by both topic models. Let $\mathbf{t}^1_{i}$ be the term distribution of the $i$-th topic of the first topic model and $\mathbf{t}^2_{j}$ be the term distribution of the $j$-th topic of the first topic model. We can then compute the pairwise cosine similarity of $\mathbf{t}^1_{i}$ to all topics $\mathbf{t}^2_{j}$. If we normalize by the sum of the pairwise cosine similarities, we obtain a probability vector $\mathbf{s}^{12}_i=(S_{i1},\ldots,S_{ik})$ with
$$    S_{ij} = \frac{S(\mathbf{t}^1_{i},\mathbf{t}^2_{j})}{\sum_{l=1}^k S(\mathbf{t}^1_{i},\mathbf{t}^2_{l})} \ .
$$
If there is no uncertainty, then one entry of $\mathbf{s}_i$ would be $1$ and all others $0$. Without loss of generality, we assume $S_{i1}=1$ and have no uncertainty for vector  $\mathbf{s}_{\min}=(1,0,\ldots,0)$.
On the other hand, we would obtain highest uncertainty, if $\mathbf{t}^1_{i}$ is equally similar to all topics $\mathbf{t}^2_{j}$ of the second topic model, i.e., no match can be found. Highest uncertainty is therefore represented by vector  $\mathbf{s}_{\max}=(\frac1k,\ldots,\frac1k)$.
The amount of matching uncertainty can, therefore, be computed by adopting the uncertainty measure by Al-Taie et al.~\cite{Al-Taie2014} based on the Kullback-Leibler divergence (Eq.~\ref{eq:KL}) as
$$
    U_M^{12}(i) = 1 - \frac{D_{KL}(\mathbf{s}^{12}_i \| \mathbf{s}_{\max})}{D_{KL}(\mathbf{s}_{min} \| \mathbf{s}_{\max})} \ .
$$
Finally, to estimate how certain a topic is, we should match it against all other topic models. Given a topic-modeling ensemble $E$, we define the matching uncertainty of one topic as the average of the values when matched against all other topic models in the ensemble:
\begin{equation}
    U_M(i) = \frac{1}{|E|-1}\sum_{l=2}^{|E|}U_M^{1l}(i) \ .
        \label{eq:UM}
\end{equation}

As a second measure we introduce \textit{existence uncertainty}. It shall measure how (un-)certain it is that a detected topic is really a topic of the given corpus. Here, a topic is considered certain, if a similar topic exists in all other topic models of the ensemble. Assume we consider again the $i$-th topic of the first topic model. Then, we only take into account the maximum similarity 
of the considered topic to all topics of every other topic model. Consequently, we define existence uncertainty as
\begin{equation}
U_E(i) = 1 - \frac{1}{|E|-1} \sum_{l=2}^{|E|}\max_{j=1}^k S(\mathbf{t}^1_{i},\mathbf{t}^l_{j}) \ .
        \label{eq:UE}
\end{equation}

\section{Uncertainty-aware Visualization}
\label{sec:visualization}



\medskip
\noindent
\textbf{Requirement Analysis.} Since our goal was to develop an interactive visual analysis tool for topic models, there are classical analysis goals that our tool should support. Those can be summarized as:
    (R1) Relating topics to terms: What is the term distribution of the detected topics?
    (R2) Relating topics to documents: What are the documents that contribute to the topics?
    (R3) Relating documents to topics: What is the topic distribution of the documents?

The second group of requirements involve the uncertainty awareness during the analysis and ensemble visualizations. We collaboratively identified the following requirements:
    (R4) Relating uncertainty to topics: What is the uncertainty in the topics?
    (R5) Relating uncertainty to ensemble: How big is the uncertainty in the ensemble?
    (R6) Relating topics to ensemble: Which topics exist in all topic models of the ensemble?
    (R7) Relating topics to topics: How similar are topics of different topic models to each other?


\medskip
\noindent
\textbf{Design Choices.} The requirements formulated above investigate different facets of the topic-modeling ensemble at different levels of abstraction. Hence, we also need to consider different visual encodings that represent operate on the different levels and can be jointly interacted with using coordinated views.

\medskip
\noindent
\textsc{Ensemble View.} We first want to provide an \textit{overview} over all topics of the ensemble. Such a view shall allow to relate topics to each other (R7). This is basically a multi-dimensional data visualization problem, where the objects are the topics and the dimensions are term probabilities for all given terms. As each topic of each topic model is an own object, we have $k\cdot |E|$ objects. Thus, we should use a method that scales well with the number of objects. Point-based visual encodings (scatterplots or scatterplot matrices) 
 are preferred over line-based techniques (parallel coordinates 
 or star glyphs) 
 and table-based techniques (heatmaps 
 or Table Lens). 
Moreover, the number of terms may also be high such that we also should use a method that scales well with the number of dimensions. Thus, scatterplots generated by dimensionality reduction shall be preferred over scatterplot matrices. The drawback of using dimensionality reduction is the loss of information. However, since the overview shall be the starting point for more detailed analysis, visual scalability is more important than loss of information at this level of abstraction. 

Many dimensionality reduction methods exist. Linear projections such as classical metric multi-dimensional scaling (MDS) 
that minimizes the stress functional and, thus, preserves distances allow for a proper interpretation of distances. However, for our goal it is more important to observe neighborhoods of topics to interpret which topics are related to each other. We therefore chose t-distributed stochastic neighbor embedding (t-SNE)~\cite{Maaten2008}, 
which we feed with the topic similarity matrix of pairwise cosine similarities (Eq.~\ref{eq:cosine}) of all topics within the ensemble and apply the default perplexity settings of the used library~\cite{sklearnTSNE}, which was empirically determined to be appropriate.

\begin{figure}[!htb]
    \centering
    (a) \includegraphics[width=0.44\textwidth]{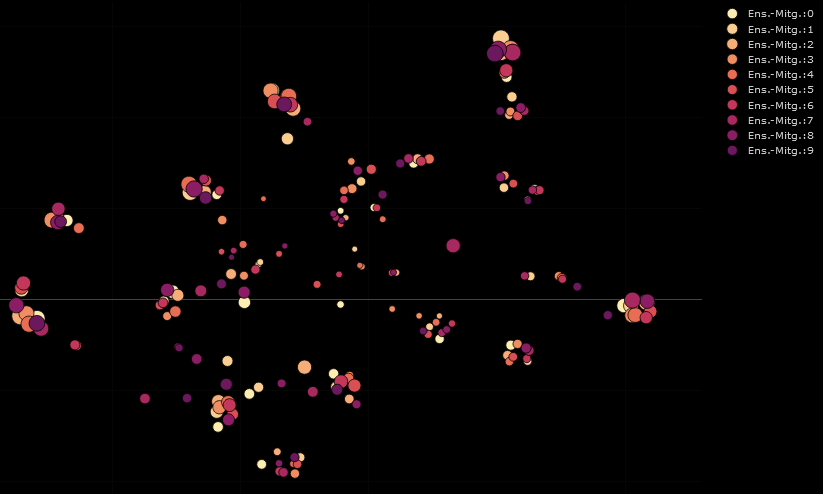} \\
    (b) \includegraphics[width=0.44\textwidth]{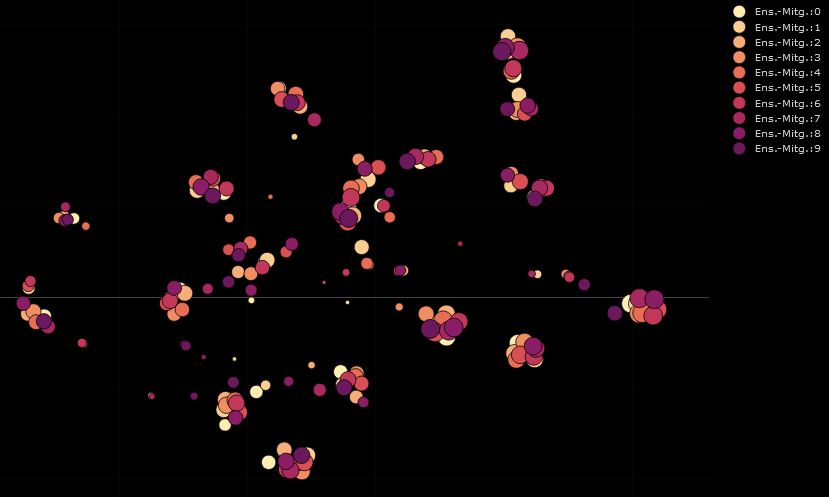} \\
    (c) \includegraphics[width=0.44\textwidth]{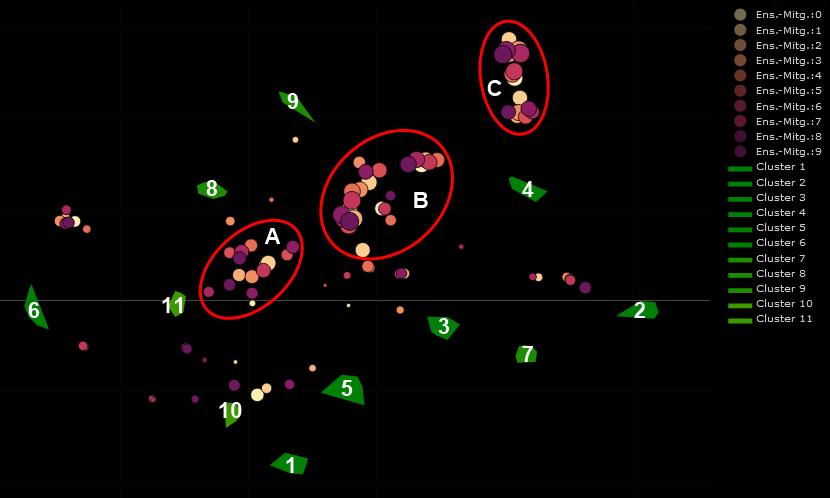}
    \caption{Ensemble overview over all topics using a t-SNE layout based on topic similarity. Topic models are encoded by color, uncertainty by size. Analyzing sampling uncertainty: Ensemble view for $E_1$ with (a) encoded matching uncertainty $U_M$, (b) encoded existence uncertainty $U_E$, and (c) grouping certain clusters with $\geq 80\%$ completeness.}
    \label{fig:ensemble1uebersicht}
\end{figure}

We next want to relate the topics to the ensemble (R6). As color is the most salient visual variable, the topic models are color-coded, where the color map can be chosen according to the ensemble type (numerical color map for model uncertainty, categorical color map for sampling uncertainty). 
We further want to visualize the uncertainty of the topics (R4) by encoding matching uncertainty $U_M$ (Eq.~\ref{eq:UM}) or existence uncertainty $U_E$ (Eq.~\ref{eq:UE}), which we encode by size. 
The more uncertain a topic is, the smaller is its rendered point. 
Figure~\ref{fig:ensemble1uebersicht}(a) provides an example for an overview visualization with topic models encoded by color and uncertainty encoded by size, which allows for judging the uncertainty in the ensemble (R5). 
Certain topics can be identified as clusters of topics from all topic models (different colors) with low uncertainties (large size), while highly uncertain topics are represented by isolated small points.



\medskip
\noindent
\textsc{Topic View.} 
To provide detailed information on the term distribution for the topics (R1) 
we support a \textit{detailed topic view}, where we assume that a few selected topics have been chosen for closer investigations. 
Methods based on dimensionality reduction are not suitable to convey exact values for term distribution. Scatterplot matrices would allow for reading off the values, but focus on pairwise correlations. What remains are line-based methods, where parallel coordinates provide a better scalability than the radial star-glyph layout, and table-based methods, where the color encoding of heatmaps provides a better scalability than the size encoding of the bars in Table Lens. Since parallel coordinates require some spacing between the axes, \textit{heatmaps} are a more compact representation that scales slightly better with the number of dimensions. 
Since the term probabilities are numerical  with same scaling, we apply one numerical color map to all dimensions.

\begin{figure}[!tb]
    \centering
    (a) {\includegraphics[width=0.44\textwidth]{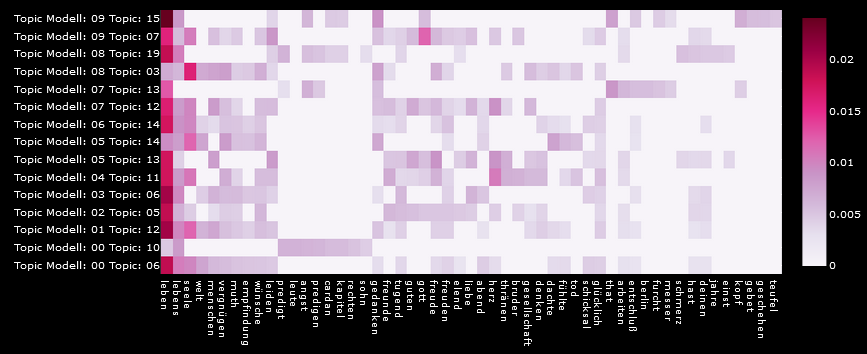}} \\
    (b) {\includegraphics[width=0.44\textwidth]{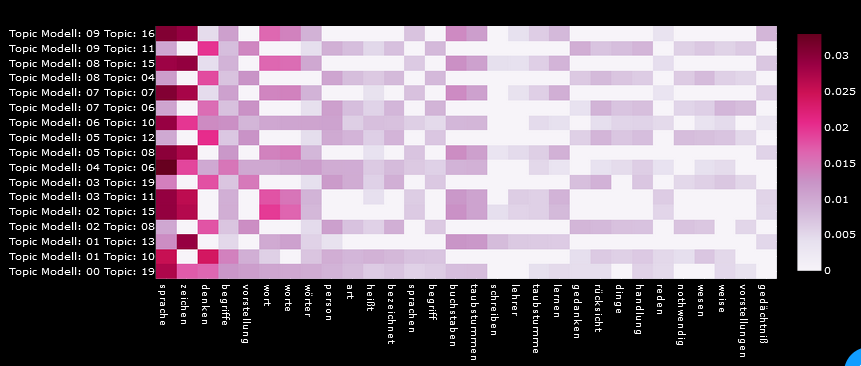}}
    \caption{Detailed topic views for $E_1$ using heatmaps for the groups (a) A and (b) 
    C identified in Figure \ref{fig:ensemble1uebersicht}(c). }
    \label{fig:ensemble1Heatmaps}
\end{figure}

For the analysis of selected topics, their most relevant terms should be investigated. Using an LDA-based topic modeling assumes that the number of relevant terms per topic is rather small. Thus, we can restrict the heatmap to only showing the set of 
most relevant terms for the selected topics. 
Figure~\ref{fig:ensemble1Heatmaps}(a) provides an example of a quite homogeneous group of topics, where the color-coded term distributions of the topics are similar. The columns are sorted decreasingly by average term relevance. The columns are labeled by the terms, the rows by the topic and the topic model. 


\begin{figure}[!tb]
    \centering
    \includegraphics[width=0.44\textwidth]{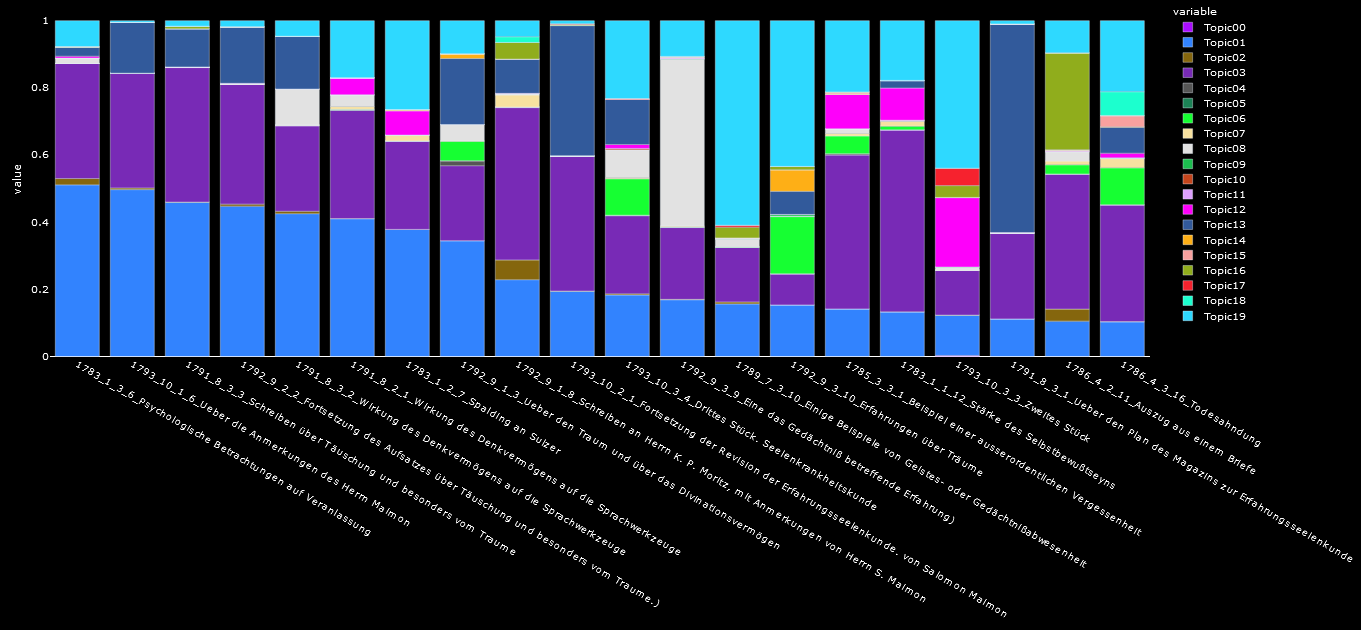}
    \caption{Topic-document view via bar chart shows the document distribution of a selected topic (blue) in descending order up to 20 documents. Stacked bar charts provide comparisons to other topics.}
    \label{fig:docs2}
\end{figure}

\medskip
\noindent
\textsc{Topic-document View.} 
To provide information about which documents contribute to the topics (R2), we visualize 
\textit{document distribution of topics}. 
Here, (stacked) \textit{bar charts} are commonly used, as they are simple and effective to provide the respective quantitative information. Figure~\ref{fig:docs2} shows the bar chart of the 20 strongest contributions of a selected topic (blue) to the documents of the corpus in descending order. The stacked bar chart shows all other topics of the topic model in comparison. 
As the topics have no order, a categorical color map is used for the topics as provided in the legend.

\begin{figure}[!h]
    \centering
    \includegraphics[width=\columnwidth]{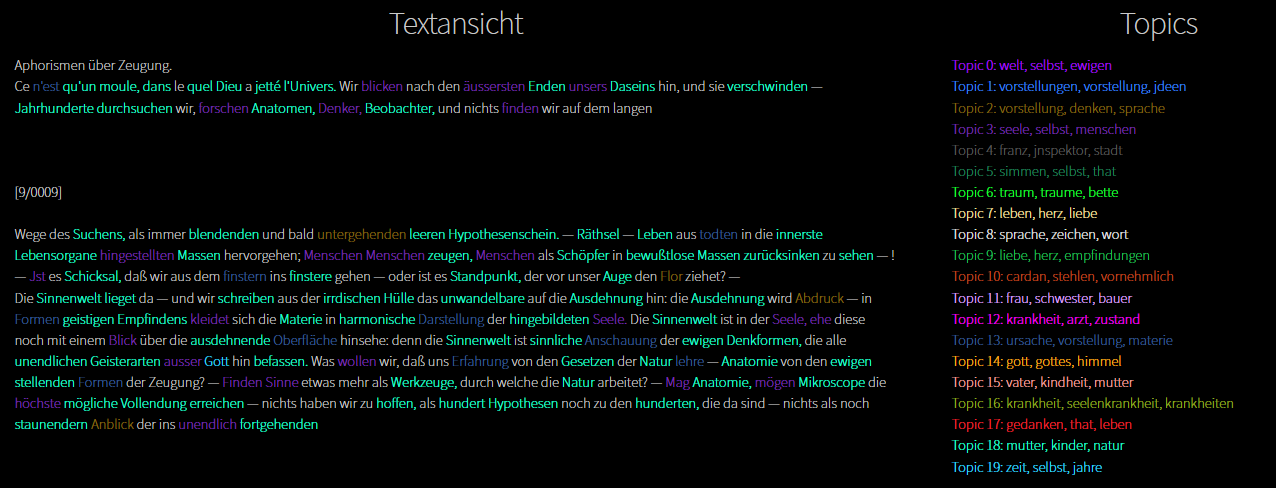}
    \caption{Close reading is supported by the document view, where the terms are highlighted in the color of the corresponding topics.}
    \label{fig:text}
\end{figure}

\medskip
\noindent
\textsc{Document View.} Finally, the provided distant-reading methods should be complemented with \textit{close-reading approaches}. Thus, the texts of individual documents are provided on demand, which allows for relating the document to the topics (R3). 
All terms are highlighted in the color of the topic with highest probability, see Figure~\ref{fig:text} using the same colors as in  the topic-document view (cf.~Figure~\ref{fig:docs2}).

\medskip
\noindent
\textbf{Uncertainty-aware Analysis Tool.}
The presented views visualize different facets of the topic modeling ensemble and jointly fulfill the formulated requirements. They are embedded in an interactive visual analysis tool that has been implemented as a web application in a platform-independent fashion using Python~\cite{python} and Dash~\cite{dash}. As input it expects the topic modeling ensemble data that is output by MALLET~\cite{mallet}. For close reading, the respective documents of the corpus need to be loaded to our tool. 

A typical analytical workflow would start with the ensemble overview (cf.~Figure~\ref{fig:ensemble1uebersicht}). 
Following the concept of \textit{showing something conditionally}~\cite{yi07} we provide filtered views: 
(1) A simple filter  restricts the topics to those that are interactively selected within the ensemble view. (2) We can filter by terms, i.e., we only show those topics, where the selected term(s) are among the most important ones. 
(3) We can filter by uncertainty, i.e., by applying an uncertainty threshold for matching uncertainty $U_M$ (Eq.~\ref{eq:UM}) or existence uncertainty $U_E$ (Eq.~\ref{eq:UE}). Thus, we can focus on certain topics only, if we want to restrict the analysis only to topics that we feel comfortable to use. On the other hand, we can also focus on uncertain topics, e.g., to analyze why these topics are uncertain. 
(4) We can filter by similarity $S$ (Eq.~\ref{eq:cosine}) to a selected topic (including similarity thresholding). 
We support showing the most similar topic of each topic model, which supports investigating whether the selected topic has matching topics in all other topic models of the ensemble. 

Having selected a group of topics in the ensemble view, those can be further analyzed by using the other views. More information about topics are provided on demand in textual form, e.g., in tabular form or as pop-up labels when hovering over all views. After having analyzed a group of topics, the outcome of the analysis can also be incorporated in the ensemble view. To do so, one chooses to group the topics into one cluster and provides a label for the cluster. The cluster is then visually represented by the convex hull of the grouped points in the ensemble view, cf.~Figure~\ref{fig:ensemble1uebersicht}(c).


\section{Results and Discussion}
\label{sec:results}

We applied our methods to the uncertainty-aware visual analysis of a corpus of empirical psychology texts, which contains 352 articles collected by Karl Philipp Moritz, Karl Friedrich Pockels, and Salomon Maimon over 10 years (1783--1793). It is concerned with the development of mental disorder~\cite{fallgesch} and is considered the first magazine in psychology in Germany~\cite{MeZ}.
The corpus was analyzed by Sieg~\cite{SiegArbeit} 
using topic modeling without taking into account uncertainties.
More precisely, 20 topics of a single topic model after fine-tuning the parameters were interpreted in a literary study.
The study also used MALLET for LDA-based topic modeling after several pre-processing steps including stop word removal and stemming. 
We used the same set-up for creating our results.

In our uncertainty-aware analysis, we will discuss the \textit{completeness of topic clusters} and the \textit{stability of topics}.
A topic cluster is considered to be complete, if it contains a topic of each topic model within the ensemble.
A topic is considered stable, if its uncertainty is below a threshold, where we empirically identified an uncertainty threshold of 0.3 to be meaningful.
On the other hand, a topic is considered unstable, if its uncertainty is above a threshold, where we empirically identified an uncertainty threshold of 0.5 to be meaningful.
We deliberately decided to leave a grey area of topics that are  considered neither stable nor unstable.


\medskip
\noindent
\textbf{Sampling Uncertainty.} To capture sampling uncertainty, we generate an ensemble $E_1$ of 10 topic models with identical parameter settings ($k=20$, $\alpha=\frac5k$, $\beta=0.01$), i.e., the ensemble comprises of 200 topics. Figure~\ref{fig:ensemble1uebersicht} shows the ensemble view with visualizing matching uncertainty $U_M$ (a) and existence uncertainty $U_E$ (b). We observe that the existence uncertainty exhibits a stronger correlation with the occurrence of clusters. 

Next, we performed a detailed analysis of all groups in the ensemble view using existence-uncertainty encoding. We identified 11 clusters with a completeness $\geq 80\%$, which we consider certain topics, cf.~green clusters Figure~\ref{fig:ensemble1uebersicht}(c). 
The terms represent some of the important terms of the corpus, e.g., Cluster 6 is about  dream experiences and Cluster 2 is about religion. 

 
 We further analyzed the three regions marked as A, B, and C in Figure~\ref{fig:ensemble1uebersicht}(c). Figure~\ref{fig:ensemble1Heatmaps} shows the respective detailed topic views using heatmaps. The heatmap of Group A shown in Figure~\ref{fig:ensemble1Heatmaps}(a) exhibits that only one term (``leben" = life/live) is strong in all topics, while other terms appear in several, but not in all. Now, it is up to the domain expert to decide whether the one term serves as an overarching theme for all topics or whether one would rather split the group into multiple subgroups that share more than only one common term. A similar observation can be made for Group B, where 
 only one term (``Seele'' = soul) is dominant in all topics. 
 Group C shown in Figure~\ref{fig:ensemble1Heatmaps}(b) is a slightly different case. Here, there are two subgroups visible that form clusters, but there are three topics that are somewhat between those two clusters. Closer inspection shows that both clusters contain topics from 7 topic models (1,2,3,5,7,8,9) while topics from 3 topic models (0,4,6) are missing. The three topics between the two cluster stem exactly from those 3 topic models (0,4,6). Hence, we can conclude that 7 out of 10 topic models decided to form two separate topics, while 3 out of 10 topic models decided to merge the two topics into one unifying topic. Content-wise Group C represents a discourse about deaf people. Hence, one may argue that the entire Group C forms one cluster, which then would be a $100\%$ complete cluster.

The matching uncertainty is generally higher (mean = 0.49, median = 0.53) than the existence uncertainty (mean = 0.28, median = 0.24). We computed Pearson correlation (0.14) and Spearman correlation (0.09) between the two uncertainty measures, but they do not exhibit a clear correlation. Thus, we conclude that the measures are indeed modeling different uncertainty aspects.

The existence uncertainties for the 11 clusters are between are between 0.07 and 0.3. Only one of the clusters has a slightly higher uncertainty than the average (0.28), all the others are below the average. The isolated points that were not part of a cluster or of groups A, B, and C instead exhibit an average uncertainty of 0.55, which is much higher than the average. Hence, we conclude that the existence uncertainty measure is indeed related to the formation of clusters.

\medskip
\noindent
\textbf{Hyper-parameter Optimization.} In a second experiment, we generated another ensemble $E_2$ with identical settings as for the first ensemble, but we now turn on the \textit{hyper-parameter optimization} provided by MALLET, which optimize parameters $\alpha$ and $\beta$ during the topic-modeling process. Both matching and existence uncertainty were slightly lower in the second ensemble when compared to the first ensemble. Qualitatively the outcome was similar though, i.e., there are some groups of topics that can be clustered, while other groups are less obvious and some isolated points with high uncertainties exist.

In conclusion, we observed that the randomization in the algorithm indeed introduces non-negligible uncertainties ($E_1$). Switching on the MALLET hyper-parameter optimization ($E_2$) actually achieved a reduction in the uncertainties. 

\medskip
\noindent
\textbf{Model Uncertainty for Hyperparameters.}
We analyzed the model uncertainty when changing parameters $\alpha$ and $\beta$ that define the properties of the Dirichlet probability distributions. We created one ensemble $E_3$ with 10 topic models by varying $\alpha$ between $0.5/k$ and $20/k$ while fixing $\beta=0.01$ and $k=20$ and another ensemble $E_4$ with 10 topic models by varying $\beta$ between $0.01$ and $0.23$ while fixing $\alpha=5/k$ and $k=20$. 
 %
 %
We, again, observe a (negative) correlation between cluster formation and existence uncertainty. For matching uncertainty, such a correlation cannot be observed. We observe a higher matching uncertainty for $E_3$ (mean and median around 0.5) when compared to $E_4$ (mean and median around 0.3), while we observe the opposite for existence uncertainty (mean and median around 0.3 for $E_3$  and above 0.4 for $E_4$). Again, we observe no correlation between the uncertainty measures and notice that stable topics with respect to one measure can be unstable with respect to the other. For matching uncertainty we observe only 36 stable (106 unstable) topics for $E_3$ and 94 stable (23 unstable) topics for $E_4$, while for existence uncertainty we observe 123 stable (25 unstable) topics for $E_3$ and 75 stable (81 unstable) topics for $E_4$.
 
In conclusion, when varying parameter $\alpha$ ($E_3$), the uncertainties only slightly increased. This can be explained by the fact that $\alpha$ does actually not change the term distributions of the topics. It merely adjusts how many dominant topics a document shall have.
Parameter $\beta$ ($E_4$), however, adjusts the term distributions. Increasing $\beta$ leads to less sparse term distributions, i.e., less dominant terms, and thus higher topic similarities. Therefore, it can be explained why varying parameter $\beta$ had a larger negative impact on the uncertainties.

\medskip
\noindent
\textbf{Model Uncertainty for Number of Topics.}
In our final experiment, we generate ensemble $E_5$ by varying the number of topics $k$ from $20$ to $50$. We generate 10 topic models with $\alpha=\frac5k$ and $\beta=0.01$ leading to 293 topic in total. In the respective ensemble view with matching uncertainty (mean and median around 0.5) and existence uncertainty (mean and median around 0.3) is 
we observe a new phenomenon, namely clusters with rather low existence uncertainty.


\medskip
\noindent
\textbf{Discussion of Uncertainty Measures. } Existence uncertainty measure $U_E$ reflects which topics are likely to be real topics of the corpus and which and questionable and require a more detailed analysis. A low existence uncertainty indicate that many similar topics exist within the ensemble. Complete topic clusters comprise of topics with low existence uncertainty. The respective ensemble view allows the user to decide which topics should be interpreted as topic modeling outcome.
However, when applying $U_E$ to an ensemble with varying number of topics $k$ we have to be careful with its interpretation. For example, for ensemble $E_5$ we can detect, at most, 20 complete topic clusters, as the smallest topic model in the ensemble has $20$ topics. On the other hand, $30$ out of $50$ topics of the largest topic model will not be part of a complete topic cluster. Thus, those topics will inevitably have larger existence uncertainties. This explains the clusters with low existence uncertainty. 
Still, topics with low uncertainty can be trusted.

\begin{figure}[!h]
    \centering
    (a) {\includegraphics[width=0.44\textwidth]{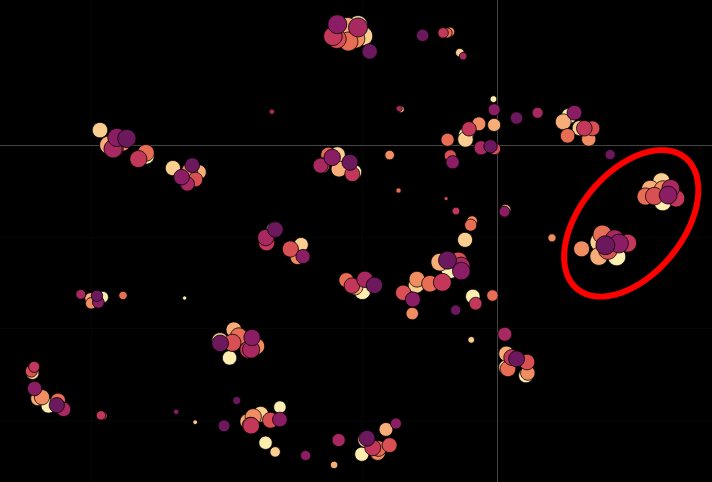}}\\
    (b) {\includegraphics[width=0.44\textwidth]{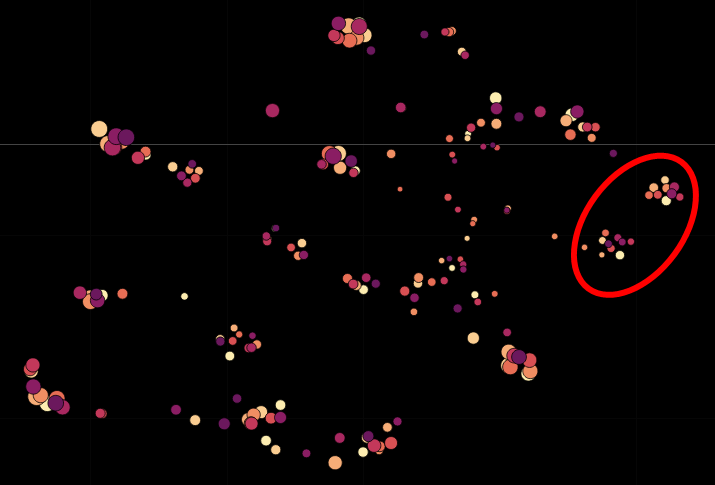}} 
    \caption{Example of two close-by topic clusters with low existence uncertainty (a), but high matching uncertainty (b). 
    }
    \label{fig:exampleUM}
\end{figure}

Matching uncertainty measure $U_M$ did not exhibit an obvious correlation with cluster formations. In particular, if two clusters get close to each other, the matching uncertainty rises. Thus, matching uncertainty can still provide helpful information. Figure~\ref{fig:exampleUM} shows an example where existence uncertainty of two close-by clusters is low (a), but matching uncertainty is high (b). Closer investigation with heatmap-based detailed topic views of the two clusters reveals slight differences, but also that the term ``Selbst'' is dominant in both clusters and other terms (such as ``Natur'' and ``finden'') also occur in both clusters. Now, the domain expert can decide whether these two clusters shall be merged or not. If doubts still occur, the expert can switch to the close-reading view of individual documents.

\medskip
\noindent
\textbf{Discussion of Visualizations.} The uncertainty-aware ensemble view allows for the interactive generation of clusters. 
The clustering is based on semantic interpretations of the domain expert. Sometimes topics are interpreted to be broader and sometimes tighter, cf.~Figure~\ref{fig:exampleUM}. Therefore, a user-centric approach is preferred over an automatic clustering. 

The topic-document and document views currently do not support ensemble visualizations. 
The scope of this paper was on the analysis of topic uncertainties acccording to (R1-R7). 
Uncertainties in document views is subject to future work.

\medskip
\noindent
\textbf{Domain Expert Feedback.} The tool was developed in close collaboration of the institutes of computer science and German philology. 
A full study as in~\cite{SiegArbeit} is surely beyond the scope of this paper 
but first interesting observations were made.
For example, when re-considering the detailed topic view in Figure~\ref{fig:ensemble1Heatmaps}(a) of Group A of Figure~\ref{fig:ensemble1uebersicht}, the domain expert noticed that while all topics share the topic (``leben'' = life/live), four topics have dominant terms related to preaching (``predigen'') and fear (``angst''), while the others have dominant terms related to desires/wishes (``wünsche''), feeling/sentiment (``empfindung''), and joy (``vergnügen''). Hence, he concludes that Group A should indeed be split accordingly to further investigate the subgroups.
The domain expert also commented that he always struggles with finding the right topic modeling settings and is never sure to have found the optimal ones.
With our tool he can observe which topics he can trust and which not (yet). 

\section{Conclusion}
\label{sec:conclusion}

We have presented an interactive visual tool for an uncertainty-aware analysis of topic modeling. Our approach is based on generating ensembles that capture sampling and model uncertainties. We proposed two measures and showed their usefulness within an application scenario. Existence uncertainty allows for judging whether a topic appears in all models, while matching uncertainty indicates clusters to be investigated for merging. Multiple views allow for overview and detail analyses of the topic-modeling ensemble. The views are integrated into a tool that addresses all facets of our requirements.

\bibliographystyle{abbrv-doi}

\bibliography{template}

\end{document}